\documentclass[12pt]{article}
\usepackage{geometry}                
\usepackage{graphicx}
\usepackage{amssymb}
\usepackage{float}
\usepackage{url,hyperref}
\title{Further comments on `Is the moon there if nobody looks? Bell inequalities and physical reality'}
\author{Richard Gill\\Mathematical Institute, Leiden University}
\date{4 May, 2023}                                           

\begin{document}
\maketitle

\begin{quote}
{\bf Abstract.} Kupczynski (2023) claims that Gill and Lambare (2022a, 2022b) misrepresent several of his published papers. This paper shows that the latest version of his ``contextuality by default'' model of a Bell experiment places no constraints whatsoever on the statistics of observed results in such experiments. It thereby effectively allows arbitrary non-locality, ie direct causal effects of local measurement settings on distant measurement outcomes. \end{quote}
\section{Introduction}
In Kupczynski (2023)\cite{MK2023}, the author (referred to as MK in the sequel) summarizes his work, and in particular, the results of Kupczynski (2020)\cite{MK2020}, in two equations labelled (1) and (3). These are two different models for Bell experiments, and it seems that over recent years MK is slowly transitioning from model (1) to model (3). Here we concentrate on model (3). 

We will use $x$ and $y$ for outcomes of measurements, taken to be elements of the set $\{-1, +1\}$, while $a$ and $b$ will stand for measurement settings, elements of the set $\{1, 2\}$. In each trial of a Bell experiment, settings $a$ and $b$ are input into two distant devices, and shortly thereafter outcomes $x$ and $y$ are output by the two devices. This is repeated many times. We consider the whole experiment as producing a long sequence of repetitions of observations of a random vector $(A, B, X, Y)$. Writing $(X_{ab},Y_{ab})$ for a pair of random variables with the joint distribution of $(X, Y)$ conditional on $A = a$, $B = b$, MK's model (3) states 
$$E(X_{ab}Y_{ab}) ~=~ \sum_{\lambda \in \Lambda_{ab}} X_a(\lambda_1, \lambda_a) Y_b(\lambda_2, \lambda_b) p(\lambda_1, \lambda_2) p_{ab}(\lambda_a, \lambda_b)\eqno(3)$$

\noindent where $\Lambda_{ab}$ is a set of four-tuples $(\lambda_1, \lambda_2, \lambda_a, \lambda_b)$.
Notice that neither of the two probability mass functions $p(\lambda_1, \lambda_2)$ and $p_{ab}(\lambda_a, \lambda_b)$ are assumed to factor. MK says ``it is clear that neither the Gill-Lambare probabilistic model nor Bell averaging over instrument variables may be used to prove CHSH inequalities for random experiments described by this probabilistic model''. As we will show, things are even worse. Model (3) is effectively no model at all. It allows anything, including serious non-locality.

As a side remark, MK's model (1) is defined through a similar equation for $E(XY | A = a, B = b, XY \ne 0)$ in which $p_{ab}$ does factor and the outcomes are elements of the set $\{-1, 0, +1\}$. Mathematically, that is the old detection loophole model, and we do not discuss it here.

In the paper we are discussing MK does not mention one of the key background assumptions which he makes in many earlier papers. That assumption is that the sets within which the hidden variables $\lambda_1$, $\lambda_2$, $\lambda_a$ and $\lambda_b$ vary are all disjoint. Now, if $\lambda_a$ takes values in a set $\Lambda_a$, and $\lambda_b$ takes values in a set $\Lambda_b$, and those two sets are disjoint, then an element $\lambda \in \Lambda_a \cup \Lambda_b$ lies in one and only one of the two sets $\Lambda_a$, $\Lambda_b$. An actual value $\lambda_a$ or $\lambda_b$ determines the value of the setting $a$ and moreover whether it belongs to Alice's or to Bob's apparatus. The actually used setting, as well as the user of the setting, is a function of the instrument setting-dependent hidden variable.

This seems strange. Typical local hidden variables models, including contextual local hidden variables models, are constructed with hidden variables taking values in $\mathbb R^p$ for some small number $p$; one will typically use the same Euclidean space independently of the setting and independently of which wing of the experiment we are talking about. Similarly, the two source hidden variables are likely taken to be elements of the same set. But one can always artificially make such spaces disjoint: for the instrument hidden variables, replace $\mathbb R^p$ with its Cartesian product with a one-point set containing the ordered pair ``(party, setting)''. Here, ``party'' is Alice or Bob, ie, which apparatus, and ``setting'' could be an angle, or just a binary setting choice. One can carry out a similar operation on the two source hidden variables.

MK's hidden assumption of disjoint sample spaces for each local hidden variable implies that the probability density or probability mass function which he sometimes calls $p_{ab}(\lambda_a, \lambda_b)$ and sometimes calls just $p(\lambda_a, \lambda_b)$ is indeed a whole \emph{family} of probability mass functions, each one defined on a different set $\Lambda_a\times\Lambda_b$. The subscript ``$ab$'' is actually superfluous since implied by the arguments.

The notation is ambiguous: the $a$ in $\lambda_a$ could take the values 1 or 2, but $\lambda_{a|a = 1}$ is not $\lambda_1$. MK tries to increase clarity by using the short-hand notation, common in both applied statistics and in theoretical physics, of writing ``$p$'' for a generic mass function; which one is meant is indicated by the \emph{name} of the variable at which it is evaluated.  

Anyway, MK wants to allow the pairs $\lambda_a, \lambda_b$ to be statistically dependent, where moreover Alice's $a$ is a function of $\lambda_a$ and Bob's $b$ is a function of $\lambda_b$. He states that CHSH inequalities cannot be proved from (3). Here, we can agree with him for the full 100\%. Something much more shocking is true: the model (3) allows for any arbitrary set of four correlations and marginal distributions. Any experiment generating i.i.d\ copies of some $(A, B, X, Y)$ can be described in this way; and actually, in a myriad different ways.

The proof of my assertion involves careful disambiguation of MK's notation. Perhaps there are more elegant ways to do this, but here is one which works.

Consider any four probability distributions of pairs $(X_{ab}, Y_{ab})$. I will denote their probability mass functions as $q_{ab}(x_{ab},y_{ab})$, defined on the set $\{-1, +1\}^2$. I emphasize again, these 16 probabilities are completely arbitrary subject only to the condition that they do define four probability distributions: the individual probabilities are non-negative and add up to $1$ in four groups of four.

Let $\Lambda_a$ equal $\{-1, +1\}\times\{(\textrm{``Alice''}, a)\}$ and let  $\Lambda_b$ equal $\{-1, +1\}\times\{(\textrm{``Bob''}, b)\}$ where $a, b \in \{1, 2\}$. Both sets can be considers as sets of 3-tuples consisting of an outcome $\pm 1$, a name, and a label of a setting, $1$ or $2$. Now define $p_{ab}$ on the set of six-tuples $\Lambda_a\times\Lambda_b$ by $p_{ab}(x, \textrm{``Alice''}, a, y, \textrm{``Bob''}, b) = q_{ab}(x, y)$, zero on all other points of this set; thus, it equals zero on all those points of  $\Lambda_a\times\Lambda_b$ with the third coordinate $a' \ne a$ and/or sixth coordinate $b' \ne b$. Now define $X_a(\lambda_1, \lambda_a) = x$, the first of the three coordinates of $\lambda_a$, and $Y_b(\lambda_2, \lambda_b) = y$, the first of the three coordinates of $\lambda_b$. It is superfluous to specify sample space and probability distributions for $\lambda_1$, $\lambda_2$, the hidden variables coming from the source.

This specification results in exactly the target distribution required in advance for each of the four pairs $(X_{ab}, Y_{ab})$. Since those distributions are completely arbitrary, their correlations can be anything too; there is no need whatsoever for Bell-CHSH inequalities to hold. No-signalling need not hold. MK also allows the marginal probability distribution of the settings $(A, B)$ to be arbitrary. Hence the distribution of $(A, B, X, Y)$ is arbitrary. 

I suspect that MK allows for statistically dependent settings because in the models of many of his earlier papers, and in particular, in his model (1), he employs the detection loophole, euphemistically renamed as the \emph{photon identification loophole}, see Ara\'ujo, Grangier and Larsson (2018)\cite{araujo}. Measurement outcomes lie in $\{-1, 0, +1\}$ where the outcome ``0'' means that no particle was detected. After post-selecting on detections of both particles, originally statistically independent settings may become correlated, a phenomenon already observed by Pearle (1970)\cite{pearle}.

In most of his earlier models, MK took $p_{ab}(\lambda_a, \lambda_b) = p_a(\lambda_a)p_b(\lambda_b)$. For that specification, but without allowing the detection loophole, Gill and Lambare derived Bell-CHSH inequalities, in at least three different ways. In a private communication MK has said to us ``one may postulate the existence of a probabilistic coupling, motivated by some physical/metaphysical assumptions (e.g. local realism/counterfactual definiteness), and test its plausibility''. Indeed, one may and one does. However, we (Gill and Lambare 2022a, 2022b\cite{GL2022a, GL2022b}) did not \emph{postulate} such an existence. We \emph{proved} the existence of a probabilistic coupling of the probability distributions constructed by MK himself. He started by postulating existence of various building blocks. We put them together in a different way and created a probabilistic coupling and hence could derive Bell-CHSH inequalities for the correlations in MK's original model. This worked because MK's correlations, and the correlations in our probabilistic coupling, are identical, by definition of the concept of a ``coupling''.

In Kupczynski (2023)\cite{MK2023}, MK states that his new model (3) was first put forward in his paper Kupczynski (2021)\cite{MK2021}. However, that is not quite true; in that paper MK assumes that $p_{ab}(\lambda_a, \lambda_b) = p_a(\lambda_a)p_b(\lambda_b)$. It seems to this author that the complexity of MK's notation and reasoning has led the author, over the years, deeper and deeper into misunderstanding of his own results. Each successive paper partially quotes his earlier results, but also modifies them, for instance by omitting key conditions. The mistakes possibly come about because MK does not make much use of modern probability language. He explains what he is doing by writing out long formulas for expectation values. Such formulas can be replaced by verbal descriptions using the language of random variables, probability distributions, conditional independence.  One can even go further and present graphical descriptions using the language of modern statistical causality theory based on DAGs (directed acyclic graphs), as we also do in Gill and Lambare (2022b)\cite{GL2022b}.

In conclusion, we have shown that Kupczynski's latest model of a Bell experiment places no constraints whatsoever on the statistics of the observed results. It effectively assumes non-locality of the effects of measurement settings on measurement outcomes.

\section{Some further thoughts}

MK frequently refers to spreadsheets of observations on four jointly distributed variables. It seems he is thinking of a set of observations of a quadruple of counterfactual variables $(X_1, X_2, Y_1, Y_2)$ and he says that Bell-CHSH inequalities hold for all samples from such a distribution. He is of course referring to an inequality involving the four empirical correlations between each of the $X$ variables and each of the $Y$ variables. That inequality is an elementary consequence of elementary arithmetic, and does not deserve to be called a CHSH inequality. It does feature as a lemma in a proof of the CHSH inequality. Bell experiments generate data consisting of many observations of a four-tuple $(A, B, X, Y)$. The CHSH inequalities are inequalities concerning the correlations $E(XY | A = a, B = b)$. Notice, the correlations are theoretical expectation values. The CHSH inequality follows from physical assumptions which justify the mathematical existence of a four-tuple  of counterfactual variables $(X_1, X_2, Y_1, Y_2)$, statistically independent of $(A, B))$, such that in a probabilistic coupling, $X = X_A$ and $Y = Y_B$. The counterfactuals $(X_1, X_2, Y_1, Y_2)$ are essentially the hidden variables which would exist under the hypothesis of local realism (with measurement independence).

We suspect that these MK's references to $N\times 4$ spreadsheets and finite data sets were inspired by our own paper Gill (2014)\cite{gill}, where they were used to visualise some new probabilistic results on the data from Bell experiments, when assuming local hidden variables and no time or memory loophole.

MK likes to consider local hidden variables models with measurement outcome space $\{-1, 0, +1\}$, but where the correlations studied by the experimenter are expectation values conditional on neither outcome being equal to zero. This seems to be his rationale in moving from his model (1) to his model (3). What he seems only partially to realise is that conditioning on $XY\ne 0$ when computing $E(XY|AB = ab, XY\ne 0)$ in such a context alters the joint probability distribution of \emph{all} of the variables which he postulated as somehow ``lying behind'' the originally observed variables $(A,B, X, Y)$. The joint probability density of his six hidden variables $(\lambda_1, \lambda_2, \lambda_{a|a = 1, 2}, \lambda_{b|b = 1, 2})$ and of the settings $(A, B)$, changes on conditioning on the event $XY\ne 0$. I have the impression that MK does not realise this, but sees model (3) as a consequence of model (1), after conditioning. In some sense, it certainly is: as we have explained, model (3) is \emph{always} true, whatever the distribution of $(A, B, X, Y)$. So it certainly also fits to the model obtained from (1) \emph{after} conditioning. However the individual densities in the model (3) are no longer the same as what they were in model (1), and the original independence assumptions are generally no longer true either.

\raggedright


\frenchspacing

\begin{thebibliography}{x}

\bibitem[1]{MK2023} Kupczynski, M (2023), Response: ``Commentary: Is the moon there if nobody looks? Bell inequalities and
physical reality'', Front. Phys. 11:1117843. \url{https://doi:.org10.3389/fphy.2023.1117843}

\bibitem[2]{MK2020} Kupczynski, M (2020), Is the Moon There If Nobody Looks: Bell Inequalities and Physical Reality, Front. Phys. 8:273.
\url{https://doi.org/0.3389/fphy.2020.00273}

\bibitem[3]{GL2022a} Gill, RD, and Lambare, JP (2022a), General commentary: Is the moon there if nobody looks—Bell inequalities and physical reality.
Front. Phys. 10:1024718.
\url{https://doi.org/10.3389/fphy.2022.1024718}

\bibitem[4]{GL2022b} Gill, RD, and Lambare, JP (2022b), Kupczynski's Contextual Locally Causal Probabilistic Models are constrained by Bell's theorem.
\url{https://doi.org/10.48550/arXiv.2208.09930}

\bibitem[5]{araujo} Ara\'ujo, M, Grangier, P, and Larsson, J-Å. (2018), Comment on `The photon identification loophole in EPRB experiments: computer models with single-wing selection'.
\url{https://doi.org/10.48550/arXiv.1807.04999}

\bibitem[6]{pearle} Pearle, P. (1970), Hidden-Variable Example Based upon Data Rejection, Phys. Rev. D 2, 1418-1425. \url{https://journals.aps.org/prd/abstract/10.1103/PhysRevD.2.1418}

\bibitem[7]{MK2021} Kupczynski,  M. (2021), Contextuality-by-Default Description of Bell Tests: Contextuality as the Rule and Not as an Exception. Entropy 2021, 23, 1104. 
\url{https://doi.org/10.3390/e23091104}

\bibitem[8]{gill} Gill, R.D. (2014), Statistics, Causality and Bell’s Theorem, Statist. Sci. 29(4): 512-528. \url{https:doi.org/10.1214/14-STS490}

\end{thebibliography}
\end{document}